\documentclass{article}
\usepackage{amsmath}
\usepackage{graphicx,psfrag,epsf}
\usepackage{enumerate}
\usepackage{url} 

\newcommand{\blind}{1}

\usepackage{authblk}
\usepackage{amsmath, amsfonts, amssymb}
\usepackage{algorithm}
\usepackage[noend]{algpseudocode}

\usepackage{todonotes}
\usepackage{subcaption}
\usepackage{graphicx}
\usepackage[export]{adjustbox}[2011/08/13]

\usepackage{tikz}
\usetikzlibrary{arrows.meta,shapes.geometric} 

\usepackage{hyperref}
\usepackage{float}

\newcommand{\fp}{P_{\text{fp}}}
\newcommand{\fn}{P_{\text{fn}}}

\renewcommand{\Pr}{\mathbb{P}}
\newcommand{\uti}{\Psi}
\newcommand{\esti}{\widehat{\uti}}
\newcommand{\data}{\mathbf{d}}
\newcommand{\des}{T}

\newcommand{\dess}{\des'}
\newcommand{\dataa}{\data'}

\newcommand{\pdir}{P_{\text{p}}}
\newcommand{\pin}{P_{\text{s}}}
\newcommand{\bas}{P_{\text{b}}}

\newcommand{\mpi}[1]{\Pr(\theta_{#1}=1|\des,\data)} 
\algnewcommand{\LineComment}[1]{\State \(\triangleright\) #1}

\begin{document}

\if1\blind
{

\title{\bf DOPE: D-Optimal Pooling Experimental design with
  application for SARS-CoV-2 screening}

\author[1,2]{Yair Daon\thanks{Yair Daon was supported by a
    post-doctoral fellowship from the Tel Aviv University Center for
    Combating Pandemics and the Raymond and Beverly Sackler Dean's
    Post-Doctoral fellowship. The authors would like to thank
    Profs. David M. Steinberg and Arie Tamir for valuable discussion,
    advice and support.} }

\author[1,3]{Amit Huppert}

\author[1,2]{Uri Obolski}

\affil[1]{School of Public Health, Tel Aviv University\\ Tel Aviv,
  Israel}

\affil[2]{Porter School of the Environment and Earth Sciences, Tel
  Aviv University\\ Tel Aviv, Israel}

\affil[3]{The Gertner Institute for Epidemiology and Health Policy
  Research, Tel Hashomer, Israel}

\maketitle

} \fi



\bigskip
\begin{abstract} 
Testing individuals for the presence of severe acute respiratory
syndrome coronavirus 2 (SARS-CoV-2), the pathogen causing the
coronavirus disease 2019 (COVID-19), is crucial for curtailing
transmission chains. Moreover, rapidly testing many potentially
infected individuals is often a limiting factor in controlling
COVID-19 outbreaks. Hence, \emph{pooling} strategies, wherein
individuals are grouped and tested simultaneously, are employed. We
present a novel pooling strategy that implements D-Optimal Pooling
Experimental design (DOPE). DOPE defines optimal pooled tests as those
maximizing the mutual information between data and infection
states. We estimate said mutual information via Monte-Carlo sampling
and employ a discrete optimization heuristic for maximizing it. DOPE
outperforms common pooling strategies both in terms of lower error
rates and fewer tests utilized. DOPE holds several additional
advantages: it provides posterior distributions of the probability of
infection, rather than only binary classification outcomes; it
naturally incorporates prior information of infection probabilities
and test error rates; and finally, it can be easily extended to
include other, newly discovered information regarding COVID-19. Hence,
we believe that implementation of Bayesian D-optimal experimental
design holds a great promise for the efforts of combating COVID-19 and
other future pandemics.
\end{abstract}

\vfill


\section{Introduction}
During the current COVID-19 pandemic, large-scale testing efforts for
detecting the presence of the SARS-CoV-2 virus, the causative agent of
the disease, are crucial. Testing allows isolating infected
individuals, thus breaking transmission chains. Testing for SARS-CoV-2
is typically done using RT-PCR (reverse transcriptase polymerase chain
reaction, see Section~\ref{subsec:likelihood} for details). Testing
via RT-PCR kits can be a limiting factor, thus creating a bottleneck
in screening and isolation efforts \cite{MinaNEJM, MinaScience}. The
most common way to increase efficiency and throughput of RT-PCR tests
is \emph{pooling}. Pooling is the act of using samples from several
different individuals in one RT-PCR test, hereby referred to as a
\emph{pool}. Several pooling strategies have been previously suggested
\cite{DorfmanOriginal, CompressedPooling}, analyzed \cite{Kim,
  OptimalDorfmanPool}, and applied \cite{DorfmanYuvalDor,
  NaturePooling, MatrixDorfman, RecursiveSevenFold}. The modus
operandi of pooling is as follows: A result is observed for one or
several pools, and then further action is taken. Usually, a negative
result for a pool means all members of said pool are declared negative
without any further testing. A positive result, on the other hand, may
render some individuals positive or require further testing.

\subsection{Pooling strategies}\label{subsec:strategies}
Pooling originated in the seminal work of Dorfman
\cite{DorfmanOriginal} in 1943. Since then, pooling has evolved into
what is known today as group testing \cite{GroupTestingSurvey}. There
are several common pooling strategies, and they are outlined
below. Implementation details can be found in Supplementary Material
A.

\emph{Dorfman} pooling \cite{DorfmanOriginal} starts by testing a
predetermined number of individuals in a pool. If the pooled test
result is negative, all pool members are declared negative. Otherwise,
each one is tested separately. A large scale testing effort
\cite{DorfmanYuvalDor} has shown that Dorfman pooling can save 76\% of
RT-PCR tests.

In \emph{recursive} pooling \cite{Kim}, if the first pooled test is
positive, the pool is split into two and the process
repeats. Otherwise, all pool members are declared negative. Thus, an
individual is only declared positive if they are eventually tested
separately and the test result is positive. One study showed a
recursive pooling can potentially result in a seven-fold increase in
throughput \cite{RecursiveSevenFold}.

\emph{Matrix} pooling \cite{MatrixPooling} arranges a population of
size $N=mn$ in an $m\times n$ matrix. Each row and column are then
pooled and individuals in the intersection of positive rows and
columns are tested separately. We were not able to find data of a real
world implementation of matrix pooling.


\subsection{Contribution}\label{subsec:contribution}
We develop DOPE (D-Optimal Pooling Experimental design), a novel
Bayesian pooling strategy. DOPE identifies which choice of pools
maximizes the mutual information between population infection state
and pooled test data. This choice of mutual information as an
optimization objective categorizes DOPE as a D-optimal experimental
design technique \cite{BayesianDesign} and results in superior
performance of DOPE compared to competing strategies.

DOPE is a Bayesian strategy and as such, enjoys the common advantages
of Bayesian methods. Assumptions on the population and RT-PCR test
error rates are easily incorporated into a prior and a likelihood
model, respectively. Furthermore, DOPE allows the probabilities of
infection to be naturally quantified via the posterior. These
probabilities convey more information and allow greater flexibility
compared to a binary test result.

Precise quantification of the above-mentioned probabilities of
infection allows DOPE to perform trade-offs between error rates and
number of tests as required. Most competing pooling strategies do not
allow for such an adaptive property and hence do not have control over
the number of tests or error rates.

Another advantage of DOPE is evident when considering edge cases in
competing strategies. Consider Dorfman pooling: how should one act if
the first pooled test is positive, yet all subsequent tests are
negative? Similar events arise for recursive and matrix pooling as
well, see implementation details in Supplementary Material A.
Such events all have nonnegligible probabilities under the
empirically estimated test error rates, and are likely to result in
implementation problems. In contrast, there are no ambiguous events
when DOPE is the strategy of choice. All test results are used for
updating one's beliefs via Bayes' theorem.

Lastly, DOPE is useful across both high and low infection prevalence
scenarios. Some competing strategies lose efficiency at high infection
prevalence \cite{DorfmanOriginal, Kim, OptimalDorfmanPool,
  NaturePooling}; others may suffer from increased false-negative
rates due to unmet assumptions of sparsity
\cite{CompressedPooling}. DOPE, in contrast, is inherently adaptable
and suitable for a wide range of infection prevalence levels.

\section{Methods}
DOPE is comprised of several components. Briefly, a Bayesian model for
pooling is formulated and a design is defined as a combination of
pools. An optimal design is defined as maximizing mutual information
between population infection state and pooled test data. Calculating
said mutual information proceeds via Monte-Carlo simulations. Then an
optimal design is found via discrete optimization, data are collected
and the process repeats.

\subsection{Prior}\label{subsec:prior}
The prior encodes the probability of every possible infection state of
the tested population. We assume the following structure: The
population is divided to disjoint clusters (e.g. families, work
places, classrooms), each contains a (potential) initial source of
primary infection, which occurs with probability $\pdir$. A secondary
infection of other members of the cluster occurs with probability
$\pin$ for each. If no primary infection occurs, the probability that
nonprimary members of the cluster are infected is the infection
prevalence in the general population $\bas$. Our assumptions are given
below, with their corresponding notation:

\begin{itemize}
\item Population members are denoted $\{1, \dots, N\}$.
\item The {\bf population state} is captured in $\theta \in
  \{0,1\}^{N}$. Individual $h \in \{1,\dots,N\}$ is either infected or
  not, with $\theta_h = 1$ or $\theta_h = 0$, respectively.
\item The population is partitioned into $M$ disjoint {\bf clusters}
  $C_1,\dots, C_M$. A single cluster represents, e.g., a household.
\item A {\bf cluster} $C$ is a tuple: $C=(h_0,h_1,\dots,h_n)$. We
  assume here, for sake of notation only, that all clusters contain
  the same number of members $n+1$.
\item For cluster $C$ denote $\theta^{(C)} := (\theta_{h_0},\dots,
  \theta_{h_n})$.
\item A {\bf primary} infection of $h_0$ occurs with probability
  $\pdir$.
\item A {\bf secondary} infection of any of $h_1,\dots,h_n$ by $h_0$
  occurs independently with probability $\pin$.
\item If no primary infection occurs, $h_1,\dots,h_n$ are infected
  with the {\bf basal} prevalence of infection in the general
  population $\bas$.
\end{itemize}

Since clusters are disjoint, their prior probabilities are
independent:
\begin{equation}\label{eq:prior decomposition}
\Pr(\theta) = \prod_{C\in C_1,\dots,C_M} \Pr(\theta^{(C)}).
\end{equation}

Turning our attention to cluster $C$:
\begin{align}\label{eq:cluster prior}
  \begin{split}
    \Pr(\theta^{(C)}) &= \Pr(\theta_{h_0})
    \prod_{j=1}^{n}\Pr(\theta_{h_j} | \theta_{h_0}) \\
    &= \big [\pdir \prod_{j=1}^n
      \pin^{\theta_{h_j}} (1-\pin)^{1-\theta_{h_j}} \big
    ]^{\theta_{h_0}} \big [(1-\pdir) \prod_{j=1}^n
      \bas^{\theta_{_j}}(1-\bas)^{1-\theta_{h_j}}\big
    ]^{1-\theta_{h_0}}.
  \end{split}
\end{align}

An explicit expression for $\Pr(\theta)$ is easily found from
equations \eqref{eq:prior decomposition} and \eqref{eq:cluster
  prior}. One can rightfully claim that our prior does not allow
co-infection between nonprimary household members (e.g. $h_2$ and
$h_3$). However, the difference in probabilities is negligible, see
Supplementary Material A.

\subsection{Likelihood}\label{subsec:likelihood}
The likelihood encodes our assumptions on pooled tests and how they
can err. Before delving into our probabilistic assumptions, we briefly
explain the process of testing for SARS-CoV-2 by RT-PCR. Our
exposition intentionally avoids many details and a comprehensive
review of RT-PCR can be found in \cite{CellBook}.

PCR is a process in which a targeted DNA strand's frequency is
amplified to create billions of copies in a reaction mixture. At the
end of this \emph{amplification} process, the presence of targeted DNA
molecules in a reaction mixture can be confidently
determined. Arguably, the best way to understand PCR is as a chain
reaction: Each targeted DNA molecule is copied into two, that are
copied into four and so forth. This cascade gives PCR its name:
polymerase chain reaction.

Specifically, the PCR process is comprised of cycles, with each cycle
doubling the abundance of the targeted DNA. In each cycle, each
double-strand DNA molecule is broken into two separate strands. An
enzyme called DNA polymerase generates a new double-strand DNA replica
from each separate strand. Replication cannot start without a specific
short DNA sequence, called a primer, that has to be introduced into
the reaction mixture. Introducing the right primer into the reaction
mixture ensures (almost) only the targeted DNA sequence is copied. A
protein is added to the reaction mixture, which emits light upon
successfully binding the targeted DNA. Once enough light is emitted
the tested reaction mixture is declared positive and a
\emph{detection} is said to occur. If, on the other hand, a predefined
number of cycles pass without a detection event then the reaction
mixture is declared negative.

Since the genetic material of SARS-CoV-2 is RNA rather than DNA, an
extra preprocessing step is required. An enzyme called
reverse-transcriptase (hence the RT in RT-PCR) replicates existing RNA
in the reaction mixture into DNA. Once DNA replicas are made, the PCR
process proceeds as described above, targeting SARS-CoV-2 DNA
replicas.

Failed detection of SARS-CoV-2 RNA in pooled RT-PCR testing is
referred to below as a false-negative. One possible source of
false-negatives in pooling is sample dilution. When pooling, several
samples are mixed, so the concentration of viral RNA is reduced. This
effect may cause a delay in amplification \cite{DorfmanYuvalDor}, no
detection and, consequently, a false-negative. However, \cite{Lion}
showed that this effect can be safely ignored when mixing up to 32
samples, which we correspondingly assume.

Previous studies of group testing strategies assumed that the
false-negative probability does not depend on the number of infected
samples, but merely on the existence of at least one such sample in a
pool \cite{Kim, OptimalDorfmanPool}. Current studies of pooling in the
context of SARS-CoV-2 also employ similar assumptions
\cite{Simplistic1, Simplistic2}. Specifically, these studies assume
that the probability of a negative result is the same for a pool with
a single sample coming from an infected individual and (e.g.)
five. However, in a previous study, we have shown that this assumption
does not align with experimental data \cite{CMI}. Thus, we assume that
viral RNA from each positive individual in a pool undergoes the RT-PCR
amplification process independently. Consequently, the probability of
(failed) amplification and/or detection for every sample whose source
was an infected individual is considered separately.

Erroneous detection of SARS-CoV-2 RNA (a false-positive) in pooled
RT-PCR testing can also occur. A common assumption \cite{Simplistic1,
  Simplistic2, Kim, OptimalDorfmanPool} is that the false-positive
probability does not depend on the number of negative samples in a
pool. We incorporate this assumption in our likelihood model with a
small modification. We assume that an erroneous amplification can
occur in \emph{any} pool. Specifically, it is possible that correct
amplifications fail and an erroneous one occurs simultaneously. This
assumption is relatively specific for the current application of
screening for SARS-CoV-2 via RT-PCR.  For example, cross-reactivity
with other coronaviruses would have violated this assumption, but it
was ruled out in \cite{KitComparison}.

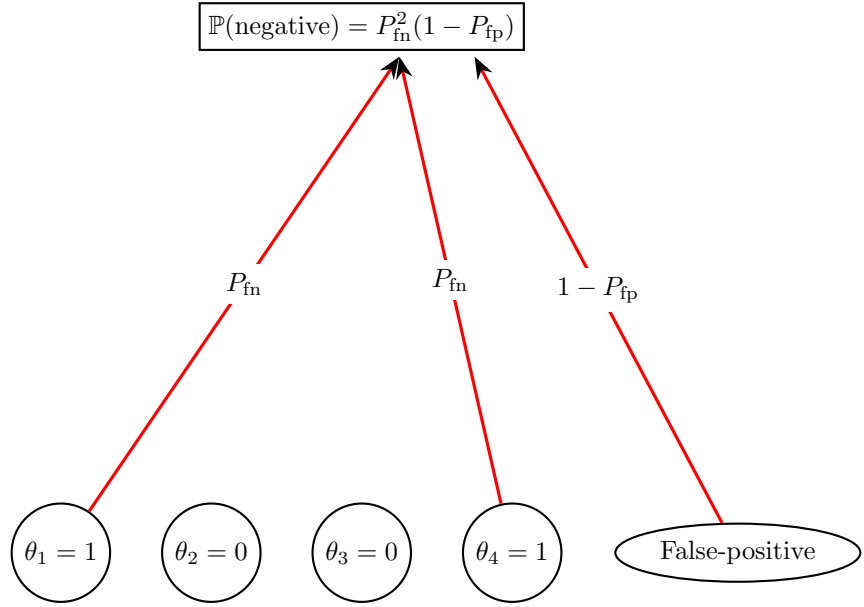
\begin{figure}[h]
  \centering
  \begin{tikzpicture}
    \begin{scope}[every node/.style={circle,thick,draw}]
      \node[shape=circle,draw=black] (A) at (0,0) {$\theta_1=1$};
      \node[shape=circle,draw=black] (B) at (2,0) {$\theta_2=0$};
      \node[shape=circle,draw=black] (D) at (4,0) {$\theta_3=0$};
      \node[shape=circle,draw=black] (E) at (6,0) {$\theta_4=1$};

      \node[shape=ellipse,draw=black]
      (F) at (9,0) {False-positive};
      
      \node[shape=rectangle,draw=black] (G) at (4,7)
           {$\Pr(\text{negative})=\fn^2(1-\fp)$};
    \end{scope}

    \begin{scope}[>={Stealth[black]},
        every node/.style={fill=white},
        every edge/.style={draw=red,very thick}]
      \path [->] (A) edge node {$\fn$} (4.5,6.6);
      \path [->] (E) edge node {$\fn$} (4.5,6.6);
      \path [->] (F) edge node {$1-\fp$} (5.5,6.6); 
    \end{scope}
  \end{tikzpicture}
  \caption{Illustration of the likelihood model of a pooled test
    result. A pool contains individuals $\{1,2,3,4\}$ with state
    $\theta=(1,0,0,1)$ (i.e. individuals 1 and 4 are infected). A
    negative pooled test implies that three detection paths failed. A
    false negative occurred for $1$ and $4$, each with probability
    $\fn$. Additionally, no erroneous detection of SARS-CoV-2
    occurred, with probability $1-\fp$. Individuals $2$ and $3$ are
    not infected and do not contribute to the probability of the
    pooled test result. A positive pooled test arises if any one of
    the above mentioned paths results in a
    detection.}\label{fig:likelihood}
\end{figure}

To summarize, we assume that for a single pool, a positive test result
is generated in one of two paths. Either, SARS-CoV-2 RNA from an
infected individual's sample is correctly amplified and detected, and
this can happen for each positive sample in a pool. Or, some erroneous
amplification occurs (e.g. contaminant viral RNA is introduced), an
event that occurs at most once per pool. Our model is illustrated in
Figure~\ref{fig:likelihood}. We now proceed to formulate the
likelihood, so we require some definitions and notations:

\begin{itemize}
\item A {\bf pool} is a collection of individuals $\{h_1, \dots, h_m\}
  \subseteq \{1,\dots,N\}$.
\item A {\bf design} $\des$ is a collection of pools. The
  $k^{\text{th}}$ pool is denoted $\des_k$.
\item Data are denoted $\data \in \{0,1\}^{|T|}$. We let $\data_k=1$
  if upon testing $\des_k$ a positive result was observed and let
  $\data_k=0$ otherwise.
\item The probability that the detection process fails for one sample
  taken from an infected individual is $\fn$.
\item The probability of an erroneous amplification and detection in a
  pooled test is $\fp$.
\end{itemize}

The probability of a negative pooled test result is presented in
\eqref{eq:likelihood a} (along with its complement
\eqref{eq:likelihood b}), and explained below.
\begin{subequations}
\begin{align}
    \Pr(\data_k=0 | \des_k, \theta) &= (1-\fp)\prod_{h \in \des_k}
    \fn^{\theta_h}, \label{eq:likelihood a}\\
    \Pr(\data_k=1 | \des_k, \theta) &= 1 - (1-\fp)\prod_{h \in
       \des_k} \fn^{\theta_h}. \label{eq:likelihood b}
\end{align}
\end{subequations}
A negative pooled test result occurs when all detection paths (both
correct and erroneous) fail. The probability of no false-detection
accounts for the $1-\fp$ term. The probability of no correct detection
is $\fn$ per infected individual. The probability that all such paths
fail is the product of the above mentioned terms, displayed in
\eqref{eq:likelihood a}. The probability of a positive result,
presented in \eqref{eq:likelihood b}, is simply the complement.
Combining \eqref{eq:likelihood a} and \eqref{eq:likelihood b}, and
recalling that $\data_k \in \{0,1\}$ yields:

\begin{align}\label{eq:combined likelihood}
  \begin{split}
    \Pr(\data_k| \des_k, \theta) &= \left [(1-\fp)\prod_{h \in \des_k}
      \fn^{\theta_h} \right ]^{1-\data_k} \left [1 - (1-\fp)\prod_{h
        \in \des_k} \fn^{\theta_h} \right ]^{\data_k}.
  \end{split}
\end{align}

Since different tests are assumed independent, the full likelihood is
the product:

\begin{align}\label{eq:likelihood}
  \begin{split}
    \Pr(\data | \des, \theta)
    &= \prod_{k=1}^{|T|} \Pr(\data_k|\des_k,\theta).
  \end{split}
\end{align}

\subsection{D-optimal design}\label{subsec:utility}
In Bayesian experimental design, a design is called \emph{D-optimal}
if it maximizes any one of several equivalent information theoretic
design criteria \cite{Marzouk,BayesianDesign,OriginalDOptimal,
  BiasedEstimator}. For convenience, we consider the mutual
information between parameters and data as the optimization
criterion. For a given design $\des$, the mutual information between
data $\data$ and population infection state $\theta$ is denoted
$\uti(\des)$:

\begin{equation}\label{eq:utility}
\uti(\des):= I(\theta;\data|\des) = \sum_{\theta,\data}
\Pr(\theta,\data|\des) \log
\frac{\Pr(\theta,\data|\des)}{\Pr(\theta)\Pr(\data|\des)}.
\end{equation}

It is known that maximizing mutual information is equivalent to
minimizing the expected posterior entropy and maximizing expected
relative entropy between posterior and prior \cite{BayesianDesign,
  Marzouk, VariationalOptimalDesign, BiasedEstimator}. Some details of
the D-optimal approach are discussed in Section
\ref{section:discussion}.

There is no closed form expression for $\uti$ and we estimate it via
Monte-Carlo sampling. We start with a straightforward calculation:
\begin{align}\label{eq:info gain}
  \begin{split}
    \uti(\des) &= \sum_{\theta,\data} \Pr(\theta,\data|\des) \log
    \frac{\Pr(\theta,\data|\des)}{\Pr(\theta)\Pr(\data|\des)}\\
    &= \sum_{\theta,\data} \Pr(\theta,\data|\des)
    \log\frac{\Pr(\data|\theta,\des)\Pr(\theta)}{\Pr(\theta)\Pr(\data|\des)}\\
    &=\sum_{\theta,\data} \Pr(\theta,\data|\des) \Big( \log
    \Pr(\data|\theta,\des) -\log \Pr(\data|\des) \Big).
  \end{split}
\end{align}
Estimating the last sum requires three steps
\cite{BiasedEstimator,Marzouk}:
\begin{enumerate}
\item Sample $\Pr(\theta,\data|\des)$.
\item Evaluate $\log \Pr(\data|\theta,\des)$ and estimate $\log
  \Pr(\data|\des)$ for each sample.
\item Average.
\end{enumerate}

We carry out the first step by sampling the prior $L$ times: $\eta_k
\sim \Pr(\theta), k=1,\dots,L$. Once we obtain the prior samples
$\eta_k$, the likelihood is sampled $Y_k \sim \Pr(\data|\eta_k,\des),
k=1,\dots,L$. This procedure results in $L$ pairs of samples from the
joint distribution of states and data: $(\eta_k, Y_k) \sim \Pr(\theta,
\data|\des)$.

Calculating the left summand $\log \Pr(Y_k|\eta_k,\des)$ is
straightforward and only requires evaluating the likelihood. The right
summand satisfies:
\begin{equation}\label{eq:precursor}
    \Pr(Y_k|T) = \mathbb{E}_{\theta} \Big[\Pr(Y_k,\theta|\des) \Big]
    =\sum_{\theta} \Pr(\theta)\Pr(Y_k|\theta,\des),
\end{equation}
and we estimate it via Monte-Carlo, taking advantage of existing
samples:
\begin{equation}\label{eq:evidence}
  \widehat{\Pr(Y_k|T)} := \frac{1}{L} \sum_{r=1}^L
    \Pr(Y_k|\eta_r,\des).
  \end{equation}

The third step is realized by first utilizing the samples $\eta_k,
Y_k$ and \eqref{eq:evidence} to define:
\begin{align}\label{eq:estimator}
  \begin{split}
    \esti(\des) &:= \frac{1}{L} \sum_{k=1}^L \Big (\log
    \Pr(Y_k|\eta_k,\des) -\log \widehat{\Pr(Y_k|\des)} \Big ) \\
    &= \frac{1}{L} \sum_{k=1}^L \Big (\log
    \Pr(Y_k|\eta_k,\des) -\log \frac{1}{L}\sum_{r=1}^L \Pr(Y_k|\eta_r,
    \des) \Big ).
   \end{split}
\end{align}

Calculating $\esti$ via equation \eqref{eq:estimator} constitutes one
of the main computational difficulties in finding an optimal
design. The logic is that the number of likelihood evaluations is
$L^2$, so calculating $\esti$ is $\mathcal{O}(L^2)$. The estimator
$\esti$ is biased and its bias is $\mathcal{O}(L^{-1})$. See
\cite{BiasedEstimator, Marzouk} for a full discussion of convergence
and bias of $\esti$. See Supplementary Material A for a discussion of
the choice of number of samples $L$.

\subsection{Posterior}\label{subsec:gibbs}
Once data $\dataa$ for design $\dess$ have been observed, we would
like to define $\uti$ for a new design $\des$. The definition is a
natural extension of \eqref{eq:utility}, with the posterior
$\Pr(\theta|\dess,\dataa)$ taking the place of the prior
$\Pr(\theta)$:

\begin{equation}
  \uti(\des;\dess,\dataa) := I(\theta,\data|\dess,\dataa) = \sum_{\theta, \data} \Pr(\theta,\data|\dess,
  \dataa, \des) \log \frac{\Pr(\theta,\data|\dess,\dataa,\des)}{\Pr(\data|\dess,\dataa,\des)\Pr(\theta|\dess,\dataa)},
\end{equation}
where $\data$ is the data for $\des$. Before data are observed and
design generated, we write $\dess =\varnothing$ and $\dataa =
\varnothing$. Therefore, $\Pr(\theta) = \Pr(\theta| \varnothing,
\varnothing)$ and indeed $\uti(\des) =
\uti(\des;\varnothing,\varnothing)$.

The calculation of $\esti(\des;\dess,\dataa)$ proceeds verbatim as in
Section~\ref{subsec:utility}. The only difference is that instead of
sampling $\eta_k \sim \Pr(\theta),k=1,\dots,L$, we sample from the
posterior $\eta_k \sim \Pr(\theta|\dataa,\dess),k=1,\dots,L$.

Sampling the posterior is achieved by Gibbs sampling. Denote all
$\theta_j$'s except the $i^{\text{th}}$ by $\theta_{-i} = \{\theta_1,
\dots, \theta_{i-1}, \theta_{i+1}, \dots, \theta_N\}$. Gibbs sampling
requires repeatedly sampling from $\Pr(\theta_i|\dess, \dataa,
\theta_{-i})$, which are calculated as follows:

\begin{equation}
    \Pr(\theta_i| \dess, \dataa, \theta_{-i}) = \frac{\Pr(\theta_i,
      \theta_{-i} |\dess, \dataa)}{\Pr(\theta_{-i}|\dess, \dataa)} =
    \frac{\Pr(\theta_i, \theta_{-i} |\dess, \dataa)}{\sum_{x\in\{0,1\}}
      \Pr(\theta_i=x, \theta_{-i}|\dess, \dataa)},
\end{equation}
and the normalization constant cancels out, making the calculation
possible.

Naively utilizing samples from the Gibbs sampler for calculating
$\esti(\des;\dess,\dataa)$ is wasteful. Recall, from the discussion in
Section~\ref{subsec:utility}, that said calculation is
$\mathcal{O}(L^2)$, where $L$ is the number of Monte-Carlo samples
utilized. Since Gibbs sampler does not generate independent samples,
naively taking $L$ samples from the Gibbs sampler would require huge
$L$ to cover all state space for $\theta$, thus rendering the
calculation of $\esti$ prohibitively expensive. A remedy is found in
\cite{Sokal}: ``The number of 'effectively independent samples' in a
run of length $n$ is roughly $n/(2\tau_{int,f})$'', where
$\tau_{int,f}$ is the integrated autocorrelation time for function
$f$. Thus, we first estimate $\tau_{int,f_i}$ for the coordinate
projections $f_i(\theta) = \theta_i$ and take $\tau :=
\max_i\tau_{int,f_i}$. The calculation of $\tau_{int,f_i}$ is carried
out using emcee's \cite{Emcee} method \texttt{autocorr} during the
chain's burn-in time. We then run the Gibbs sampler for $\tau L$ steps
and discard all but every $\tau^{\text{th}}$ sample, thus keeping
computational costs and variance for $\esti$ low. Pseudocode for our
Gibbs sampler can be found in Supplementary Material A.


\subsection{Optimization}\label{subsec:optimization}
Given a routine that calculates $\esti(\des;\dess,\dataa)$ for any
design $\des$, we need to find a way to maximize $\esti$ over all
valid designs. Designs are restricted to have a fixed number of pools,
denoted $K$. Optimizing over all valid designs results in a difficult
discrete-optimization problem, which we solve via a heuristic
hill-climbing approach. Although hill-climbing is a heuristic, we have
found it to work sufficiently well. In each step, we take the current
best design and randomly perturb it several times. We then keep the
design with maximal $\esti$ as the new best design and repeat. See
Supplementary Material A for details and pseudocode.

\subsection{DOPE}\label{subsec:dope}
We now present DOPE: D-Optimal Pooling Experimental-design, summarized
in Algorithm~\ref{alg:DOPE} below. DOPE requires two parameters:
First, the number of pooled tests per step $K$. Second, a decision
interval $I\subset [0,1]$. The decision interval defines the required
certainty levels to serve as a stopping criterion for DOPE. The
meaning of $\Pr(\theta_i=1|\des,\data) \in I$ is that the state of
individual $i$ is still uncertain, so further testing is
required. DOPE stops when there is no uncertainty regarding the state
of any individual (read: $\forall i,\ \Pr(\theta_i=1|\des,\data) \not
\in I$).

DOPE typically proceeds to find $K$ optimal pools, perform the
corresponding RT-PCR tests, and repeat the process if any individual's
posterior infection probability is in $I$. However, DOPE can also be
executed in a nonsequential manner, where no retesting is
allowed. Such a nonsequential implementation can be achieved in
Algorithm~\ref{alg:DOPE} by letting $K$ be the total number of
allotted tests and $I=\varnothing$.

\begin{algorithm}
  \begin{algorithmic}[1]
    \Procedure{DOPE}{$K, I$}
    \State $\des, \data \gets \varnothing, \varnothing$

    \Repeat
    \State $\dess \gets \textproc{OptimalDesign}(K, \des,\data)$

    \State $\dataa \gets \text{PCR}(\dess)$
    \Comment{Perform RT-PCR tests for $\dess$}

    \State $\des \gets
    (\des_1,\dots,\des_{|\des|},\dess_1,\dots,\dess_{|\dess|})$
    \Comment{Concatenate}

    \State $\data \gets
    (\data_1,\dots,\data_{|\des|},\dataa_1,\dots,\dataa_{|\dess|})$
    \Comment{Concatenate}
        
    \Until{$\forall i\ \mpi{i} \not \in I$}
    \State \Return
    $\begin{bmatrix}
      \mpi{1} > 0.5\\
      \vdots \\
      \mpi{N} > 0.5
    \end{bmatrix}$
    \Comment{Classification via posterior marginals}
    \EndProcedure
  \end{algorithmic}
  \caption{DOPE: D-Optimal Pooling Experimental design.}\label{alg:DOPE}
\end{algorithm}

\subsection{Software tools}
All computations were performed using Python 3 and NumPy 1.19.4
\cite{Numpy}. Numba 0.50.1 \cite{Numba} was used to accelerate the
Gibbs sampler. Integrated autocorrelation time was calculated with
emcee's \cite{Emcee} method \texttt{autocorr}.

\section{Results}\label{section:results}
We compare DOPE to three prominent pooling strategies: Dorfman,
recursive, and matrix pooling. We present extensive simulation
results, and consider a large number of parameter choices for DOPE. We
choose $K=1$, so DOPE always finds a single optimal pool in each
step. decision intervals' lower and upper bounds are chosen from
$\{0.01,0.02,\dots,0.15\}$ and $\{0.3, 0.35,\dots, 0.95\}$,
respectively. We take a percentage nomenclature, so (e.g.) DOPE
$\alpha,\beta$ utilizes the decision interval $I=[\alpha/100,
  \beta/100]$.

There are three performance metrics with which one can evaluate
pooling strategies: false-negative rate, false-positive rate and
number of tests. We plot false-negative rates against average number
of tests and delegate plots of false-positive rates to Supplementary
Material A (the reason is explained in
Section~\ref{section:discussion}). In addition, we present the average
posterior entropy for each strategy. Although the posterior entropy is
not a performance metric per se, we choose to present it in plots. The
reason is that presenting posterior entropy shows that indeed DOPE
succeeds in maximizing this well-defined statistical criterion.

We say a pooling strategy A \emph{dominates} another strategy B for
false-negative rates if A achieves lower false-positive rates than B,
while utilizing a smaller (or equal) number of tests. Similarly, we
say that A dominates B for posterior entropy if similar conditions
apply for posterior entropy. In the results below, we show that for
both false-negative rates, as well as for posterior entropy, there 
are decision intervals for which DOPE dominates Dorfman, recursive,
and matrix pooling.

In all simulations presented in this section, we used the commonly
observed test error rates $\fn=0.2,\fp=0.01$
\cite{EstimatingRatesKucrika,EstimatingRatesLourenco, FPR}. Infection
prevalence in the tested population was realized by varying population
connectivity parameter $\pdir$ and $\pin$, which took values in
$[0.05,0.4]$ --- see Supplementary Material A for a summary of
estimates of connectivity parameters, and Supplementary Material B for
a table of simulations parameters. The basal prevalence in the
population was always set to $\bas=0.01$.

\subsection{DOPE dominates}\label{subsec:dope dominates}
\begin{figure}[h]
\centering
\includegraphics[width=\textwidth]{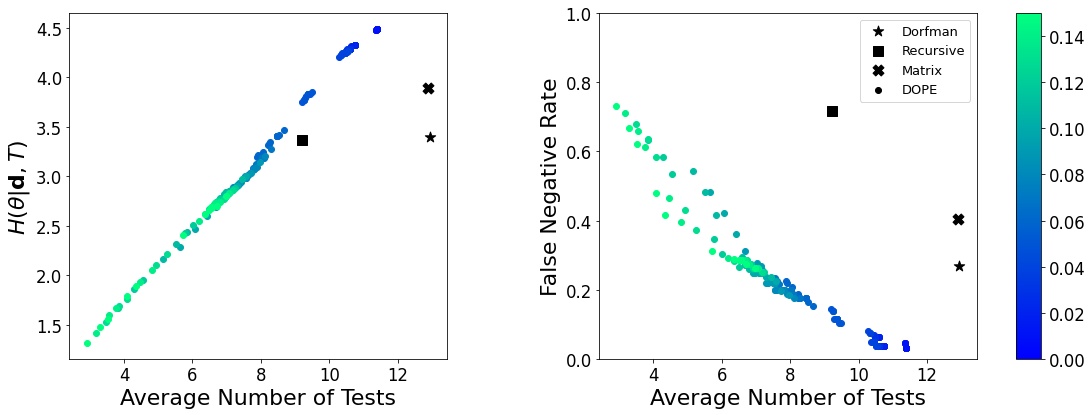}
\caption{Comparison of DOPE against other pooling strategies. We plot
  the posterior entropy (left) and false-negative rates (right)
  achieved by different pooling strategies against the number of tests
  used (x-axis). The results for DOPE ($\bullet$) are shown for
  various decision intervals (colors represent the lower bound of the
  decision interval). Dorfman ($\star$), recursive ($\blacksquare$),
  and matrix ($\mathbf{\times}$) pooling are shown for their only
  configuration, in black. We see that there are always decision
  intervals for which DOPE dominates competing
  strategies.}\label{fig:pareto big pop}
\end{figure}

We compare DOPE to Dorfman, recursive, and matrix pooling. Results are
shown in Figure~\ref{fig:pareto big pop}. We consider a population of
size $N=32$, which is the maximal pool size that can be employed
without interfering the RT-PCR process by sample dilution \cite{Lion}.
Figure~\ref{fig:pareto big pop} shows that a decision interval can be
found for which DOPE dominates competing strategies for both
false-negative rate, as well as for posterior entropy. For
presentation purposes, choices of decision intervals are grouped
according to their lower bound, with colors corresponding to the color
bar on the right of Figure~\ref{fig:pareto big pop}. E.g. DOPE 60,10,
DOPE 70,10 and DOPE 90,10 share the same color. Results are shown for
$\fn=0.2, \fp=0.01, \pdir=0.2, \pin=0.2, \bas=0.01$, which results in
average disease prevalence $\approx 7.7\%$. Number of samples in
Monte-Carlo estimation is $20000$. Number of populations simulated is
$123$.

\subsection{Varying prevalence}\label{subsec:prevalence small pop}
We examine the performance of DOPE under a wide range of disease
prevalence rates. Performance is demonstrated for a population of size
$N=10$ with infection prevalences in $[0.02,0.18]$. Connectivity
parameters generating these prevalences can be found in Supplementary
Material B. Test error rates of $\fn=0.2, \fp=0.01$ were used, with
$L=12000$ samples for the Monte-Carlo estimation.

\begin{figure}[h]
\centering
\includegraphics[width=1.2\textwidth, center]{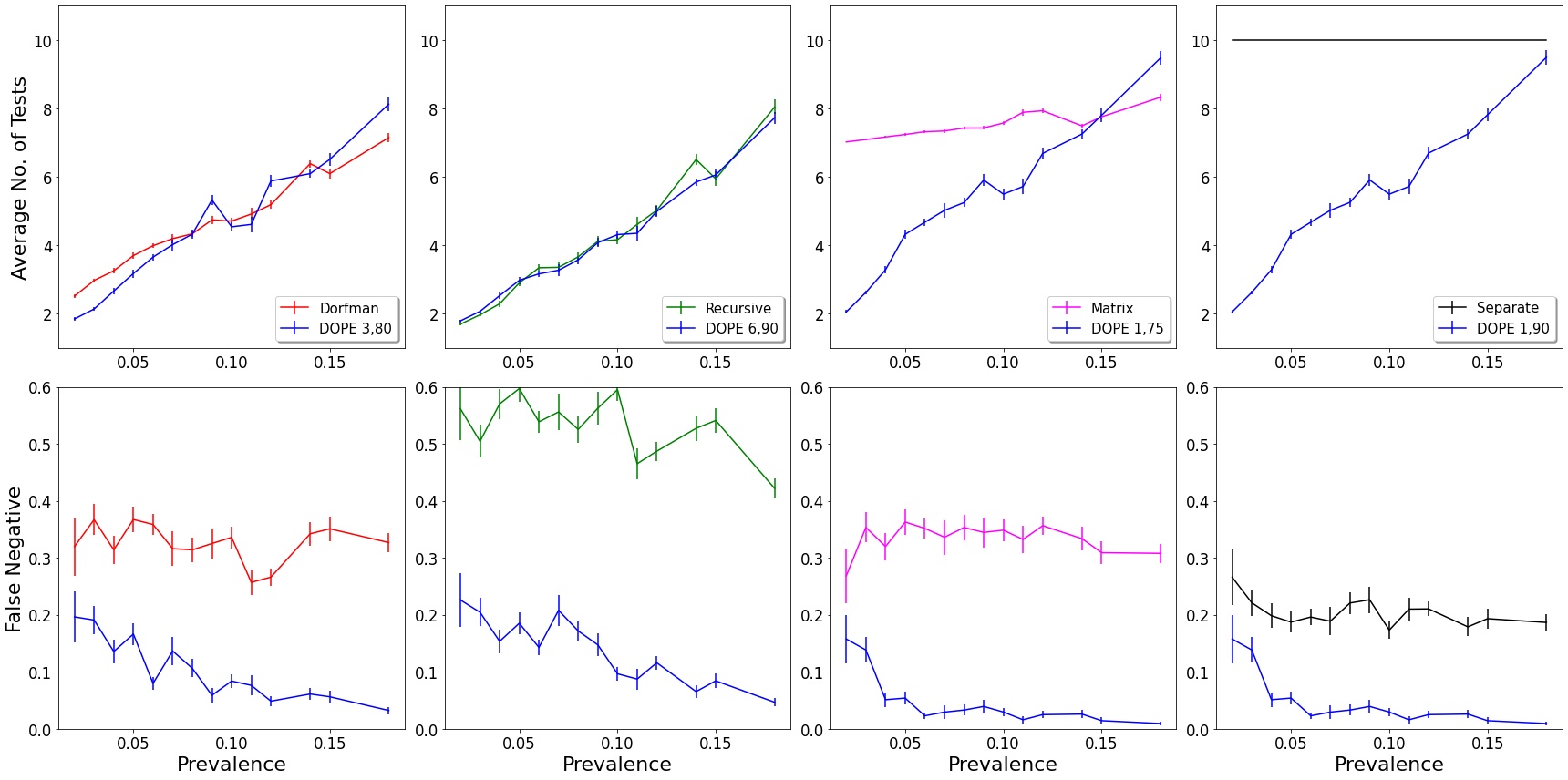}
\caption{Comparison of DOPE to other strategies under varying disease
  prevalence rates. DOPE decision intervals are chosen so that the
  number of tests taken is as close as possible to the competing
  strategies. Rows, top and bottom: number of tests and false-negative
  rate, respectively, plotted against disease prevalence. Columns show
  performance for each pair of strategies (see legend in the top
  row). DOPE, with the chosen decision intervals, consistently
  dominates each competing strategy except for matrix pooling and the
  highest prevalence.}\label{fig:prevalence small pop}
\end{figure}

In Figure~\ref{fig:prevalence small pop}, we show the performance of
DOPE for four decision intervals. Each decision interval was chosen so
that DOPE's expected number of tests was closest to one of the
competing pooling strategies (Dorfman, recursive, and matrix). We also
show such a comparison for separate testing. For each choice of
decision interval, DOPE mostly dominates the corresponding competing
strategy. The only exceptions occur when we were not able to find
decision intervals that closely match the behavior of matrix pooling
and separate testing. The reason we cannot find such decision
intervals is that the number of tests used in matrix pooling and
separate testing happens to be relatively constant in the range of
disease prevalence rates we consider. DOPE is more adaptive, hence the
number of tests it utilizes is increasing in the disease prevalence
rate. Consequently, it is difficult to find a decision interval for
which DOPE utilizes a number of tests close enough to the number of
tests utilized by either separate testing or matrix pooling throughout
the prevalence range considered.

\section{Discussion}\label{section:discussion}
In this manuscript we have presented DOPE, a novel pooling strategy
that has the potential to substantially improve the performance of
RT-PCR pooling in terms of number of tests and error rates. DOPE was
developed with the aim of maximizing the information gained from
pooled tests, precisely defined in Section~\ref{subsec:utility}. DOPE
is a Bayesian method, and as such enjoys many of the advantages
Bayesian analysis has to offer. For example, DOPE offers seamless
integration of probabilistic assumptions on population connectivity
and test errors into its underlying probabilistic model. Thus, we can
apply DOPE to tests with different error models/rates (e.g. COVID-19
antigen test \cite{AntigenWHO, AntigenCDC}). Moreover, DOPE can make
the trade-off between the number of tests and test error rates
explicit. Lastly, DOPE's error rates are lower compared to common
pooling methods, while utilizing the same amount of, or fewer,
tests. Last but no least, DOPE can return posterior infection
probabilities, giving a very refined tool for decision making under
uncertainty.

DOPE is based on an information theoretic experimental design
criterion, maximizing mutual information between population infection
state $\theta$ and pooled tests results $\data$. There are multiple
motivations for this definition. If we view the testing procedure as a
communication channel \cite{Cover}, where we wish to transmit
$\theta$, then a D-optimal design maximizes the channel capacity. The
channel capacity is the upper bound for the amount of information that
can be transmitted through the channel with vanishing error
probability, so maximizing it is sensible. Alternatively, a quick
calculation \cite{BayesianDesign} verifies that $\uti(\des) =
H(\theta) - \mathbb{E}_{\data|\des}\left [ H(\theta|\data,\des)
  \right]$, where $H(\cdot)$ is the Shannon entropy. Hence, D-optimal
designs minimize the expected posterior entropy. Since entropy is a
common measure for uncertainty, minimizing it is reasonable. Yet
another calculation \cite{Marzouk} shows that $\uti(\des)$ is the
expected relative entropy between the posterior and the
prior. Relative entropy is a common measure of ``distance'' between
probability distributions. Maximizing it roughly means we have learned
as much as possible going from prior to posterior. Our results show
that DOPE's performance is considerably superior to competing
strategies which are based on heuristics.

The Bayesian framework of DOPE also allows us to easily incorporate
test error rates into our considerations. Error rates are usually not
taken into account in the development of most pooling strategies, and
hence such strategies are not adaptive to varying error rates
\cite{Simplistic1, Simplistic2, OptimalDorfmanPool, Kim, CMI}. The
Bayesian formulation also allows DOPE to readily incorporate any prior
knowledge obtained with regards to infection probabilities of
different sub-populations. Although we have only considered
connectivity of sub-populations in this manuscript, other covariates
can potentially also be incorporated, e.g. prior data of the
likelihood of infection based on symptoms, age groups, etc.

Another important advantage of DOPE is its potential to inform
quarantine decisions in a fine-grained manner. This can be achieved by
examining DOPE's posterior infection probabilities
$\Pr(\theta|\des,\data)$, instead of its binary
classification. Utilizing this additional information, various
quarantine policies can be implemented with respect to the policy
makers' utility functions. For example, individuals with higher
posterior infection probability can be subject to a strict and
prolonged quarantine and vice versa.

By selecting appropriate decision intervals, DOPE can gauge the number
of tests it utilizes, giving rise to varying error rates, potentially
even lower than a single test's a priori error rate. We find the
required decision interval for given $\fn,\fn,\pdir,\pin,\bas$ and
cluster sizes by first simulating DOPE for many decision intervals,
e.g. $I=[\alpha, \beta]$ for $\alpha$ and $\beta$ in $\{0.01, 0.02,\dots,
0.99\}$. We then choose $I$ that utilizes the minimal number of tests
among all decision intervals that achieved error rates lower than the
desired error rate.

False-positive rates are omitted from the plots in the main text (but
are found in Supplementary Material A) since these are not the main
concern in an epidemiological context. A false-negative result has far
worse implications than a false-positive result for the spread of an
infectious disease in a susceptible population. A false-negative
implies an infected individual is not identified as such and
consequently can continue to spread the disease. In contrast, a
false-positive only implies that a noninfected individual is
unnecessarily quarantined or retested. False-positive rates are not
entirely meaningless, of course, as superfluous isolation can have
economic and social costs. However, the false-positive rates achieved
by all strategies are still very low ($\leq 1.5\%$). This is partially
because RT-PCR false-positive rates are very low to begin with
\cite{FPR}. Thus, we believe this parameter adds very little to the
comparison of competing strategies, given the vast discrepancies in
the average number of tests and false-negative rates.

DOPE, as any other strategy, has some limitations. First, epidemiological data
of population connectivity is not always available. In this case, one
can assume the population is disconnected and use this assumption as a
prior. Results for such a population are presented in Supplementary
Material A. Our simulations show that even in this case, DOPE
dominates competing strategies.

It is possible that the iterative steps required by DOPE (find optimal
pool, retest, repeat) would be difficult to implement in a real
testing scenario. In this case, a nonsequential pooling strategy,
where pools are chosen a priori and no retesting is conducted can be
implemented. We can take $K$, the number of tests per step, to equal
the number of allotted tests. Then, taking the decision interval $I
=\varnothing$ makes DOPE consequential. This is a potential future
research direction which was not in the scope of the current study.

Furthermore, DOPE requires substantial computational efforts, contrary
to Dorfman, recursive and matrix pooling that are easily
implemented. We have alleviated most computational obstacles and
currently a full DOPE run, with 10 initial starting points, population
size of $N=32$ and $L=20000$ samples takes less than five hours to run
on seven Intel(R) Xeon(R) Gold 6252 2.1GHz CPUs. For a population size
$N=10$, utilizing $L=12000$ samples, a full DOPE run takes less than a
half hour. These numbers can be considerably reduced if more CPUs are
available. Further improvements to the run time of DOPE can be
introduced. For example, it is possible that other approximations for
$\uti$, e.g. \cite{VariationalOptimalDesign}, could further reduce
DOPE's running time. Alternatively, speeding up the optimization is
also potentially possible via, e.g., solving a continuous surrogate
optimization problem. To this end, we tried employing the
$\ell_0$-sparsification method of \cite{StadlerSparsification}, but
this did not yield significant improvements in run time. Regardless,
utilizing DOPE currently requires familiarity with programming. In the
future we plan to create a GUI for easy use in facilities where
frequent testing is performed, so that large-scale use of DOPE is
possible.

Another limitation of DOPE stems from the assumptions we make. As with
all models, ours does not capture reality exactly. We neglect sample
dilution effects, and
ignore the temporal progression of the disease \cite{ViralLoadPooling}
(see the first figure of \cite{MinaNEJM} for a great illustration of
this subject).

To summarize, we have shown that Bayesian experimental design holds a
great potential for improving RT-PCR pooling. DOPE's potential to
drastically increase test throughput and decrease testing error rates
is evident. We believe further research efforts in this direction can
be very conducive to help mitigate the current pandemic, as well as
future ones.


\section{Supplementary Materials}\label{section:supp}
\begin{description}
\item[A:] Details of the Gibbs sampler and discrete optimization can
  be found in Section~A.1. A sensitivity analysis for varying
  $\fn,\fp$ is presented in Section~A.2. Simulation results for a
  disconnected population are presented in Section~A.3. Reproducing
  Figure~\ref{fig:pareto big pop} and Figure~\ref{fig:prevalence small
    pop} with false-positive data is presented in
  Section~A.4. Estimates of number of Monte-Carlo simulations is in
  Section~A.5. Some details on our implementation of competing
  strategies can be found in Section~A.6. In Section~A.7 we consider
  secondary infection probabilities, as discussed at the end of
  Section~\ref{subsec:prior}. Some parameter estimates for
  $\fn,\fp,\pdir,\pin$ are collected in Section~A.8. (pdf)
  
\item[B:] Table of parameters $\pdir, \pin$ used to generate
  populations. (csv)
\end{description}

\bibliographystyle{amsplain}
\bibliography{refs}

\end{document}